\documentclass{aa}
\usepackage{natbib,twoopt}
\usepackage[dvipsnames]{xcolor}
\bibpunct{(}{)}{;}{a}{}{,} 
\usepackage[hyphenbreaks]{breakurl}
\usepackage[breaklinks]{hyperref}      
\bibpunct{(}{)}{;}{a}{}{,}             
\definecolor{cobalt}{rgb}{0.06, 0.2, 0.65}
\hypersetup{
  colorlinks,
  citecolor=cobalt,
  linkcolor=[rgb]{0.8, 0.2, 1.0},
  urlcolor=cobalt,
}

\usepackage{txfonts}
\usepackage{newtxtext,newtxmath}
\usepackage[T1]{fontenc}

\usepackage{graphicx}	
\usepackage{amsmath}	
\usepackage{amssymb}	
\usepackage{subcaption}
\usepackage{adjustbox} 
\usepackage{tabularx}
\usepackage{upgreek}
\usepackage{orcidlink}

\newcommand{\msol}{\,M$_{\odot}$}

\newcolumntype{C}[1]{>{\centering\arraybackslash}p{#1}}

\begin{document}

\title{Investigation of mass substructure in gravitational lens system SDP\,81 with ALMA long-baseline observations}
\titlerunning{Searching for substructure in SDP\,81}

\author{
\orcidlink{0000-0002-8999-9636}H.\,R.~Stacey\inst{1,2}\thanks{E-mail: hannah.stacey@eso.org},
\orcidlink{0000-0002-4912-9943}D.\,M.~Powell\inst{2},
S.~Vegetti\inst{2},
\orcidlink{0000-0003-1787-9552}J.\,P.~McKean\inst{3,4,5},
\and
\orcidlink{0000-0003-2812-8607}D.\,Wen\inst{3}
}
\authorrunning{H.\,R.~Stacey et al.}

\institute{
European Southern Observatory (ESO), Karl-Schwarzschild Str. 2, D-85748 Garching bei M\"unchen, Germany 
\and
Max Planck Institute for Astrophysics, Karl-Schwarzschild Str. 1, D-85748 Garching bei M\"unchen, Germany
\and
Kapteyn Astronomical Institute, University of Groningen, Postbus 800, NL-9700 AV Groningen, The Netherlands
\and
South African Radio Astronomy Observatory (SARAO), P.O. Box 443, Krugersdorp 1740, South Africa
\and
Department of Physics, University of Pretoria, Lynnwood Road, Hatfield, Pretoria, 0083, South Africa
}

\date{Received 16 June 2025; accepted 3 Sept 2025}

\abstract
{The prevalence and properties of low-mass dark matter haloes serve as a crucial test for understanding the nature of dark matter, and may be constrained through the gravitational deflection of strongly lensed arcs. Previous studies found evidence for the presence of low-mass dark matter haloes in observations of the gravitationally lensed, dusty star-forming galaxy SDP\,81, using the Atacama Large Millimetre/sub-millimetre Array (ALMA). In this work, we analyse these observations to assess the robustness of these reported results. While our analysis indicates that the data support additional angular structure in the lensing mass distribution beyond an elliptical power-law density profile, we do not find evidence for two previously reported sub-halo detections. However, we verify with realistic mock data that we could have found evidence in favour of a previously reported $\approx 10^{9}\,{\rm M_{\odot}}$ sub-halo with a log Bayes factor of 29, should it exist in the real data. After testing various systematics, we find that this previous sub-halo inference was most likely spurious and resulted from an inadequate smooth model, specifically, poorly fitting multipoles. While we do not find evidence in favour of any individual sub-halo, we find evidence for similarity in the lensing signatures of multipoles ($m=3,4$) and single massive sub-haloes, consistent with other recent work. We suggest that future searches for low-mass haloes in lensed arcs include lens angular structure in the form of multipoles up to 4th order and require a good-fitting smooth model as a prerequisite. Overall, our findings demonstrate the suitability of ALMA data of this quality to simultaneously constrain the abundance of low-mass haloes and lens angular structure. }


\maketitle



\section{Introduction}

A concordance of theoretical and observational evidence suggests that most matter in the Universe is non-baryonic dark matter. The nature of dark matter is unknown, but it is understood to be fundamental to the formation and evolution of structures in the Universe (see \citealt{Wechsler:2018} for review). 

Cold dark matter (CDM) models of structure formation, in which dark matter consists of a very weakly interacting particle beyond the Standard Model, succeed in predicting large-scale structure and the global properties of galaxies at $z=0$ (see \citealt{Somerville:2015} for review). Hypothetical particle candidates include axions and neutralinos \citep{Bergstrom:2000}, which are yet undetected in ground-based experiments (e.g. \citealt{ADMXCollaboration:2021,Aalbers:2022}). However, CDM models of structure formation are largely untested on mass scales $\lesssim10^{10}$\msol. In this regime, many low-mass dark matter haloes are predicted to form and follow a well-defined mass function \citep{Moore:1999,Springel:2008,Frenk:2012} and density profile \citep{Wang:2020}. 

Possible alternatives to CDM include warm dark matter (WDM), in which the hypothetical dark matter particle has a non-negligible thermal velocity, and self-interacting dark matter (SIDM), in which the particles have a non-zero self-interaction cross-section. In contrast to CDM, WDM produces low-mass haloes that are less concentrated and less abundant \citep{Lovell:2012,Lovell:2020}, and their haloes have different density profiles \citep{Despali:2020}. On the other hand, SIDM produces haloes with more diverse properties than CDM \citep{Despali:2019,Nadler:2023}. Therefore, the abundance and properties of low-mass haloes can constrain possible dark matter particle candidates. 

In the absence of a detectable electromagnetic signature (e.g. see \citealt{Sawala:2016}), a promising way to quantify low-mass haloes at cosmological distances is through their influence on gravitationally lensed images. This is now a well-established method to constrain the shape of the halo mass function and the properties of hypothetical dark matter particles \citep{Birrer:2017,Vegetti:2012,Vegetti:2014b,Vegetti:2018,Vegetti:2024,Ritondale:2019, Gilman:2020a,Powell:2023}. Recent work suggests it may also be possible to constrain the concentration of low-mass perturbers \citep{Vegetti:2014a,Gilman:2020b,Minor:2021,Zhang:2022,Amorisco:2022,Despali:2024,Minor:2025,Enzi:2025,Tajalli:2025}. 

Here, we report our investigation of SDP\,81: a dusty star-forming galaxy at $z=3.042$ \citep{Eales:2010} that is lensed by a massive elliptical galaxy at $z=0.2999$ \citep{Negrello:2014}. Two previous studies have reported finding evidence for two different low-mass haloes in (or along the line-of-sight to) the lensing galaxy halo using Atacama Large Millimetre/sub-millimetre Array (ALMA) observations at $\approx30$\,mas angular resolution \citep{Hezaveh:2016,Inoue:2016}. Currently, \cite{Hezaveh:2016} represents the only widely cited claim of detection of a low-mass halo in interferometric data, and the third overall \citep{Vegetti:2010,Vegetti:2012}. While the two earlier reported detections in optical/infrared data have been independently verified with different methodologies \citep{Minor:2021,Sengul:2022,Nightingale:2022,Ballard:2024,Minor:2025,Despali:2024,Enzi:2025,Tajalli:2025}, the \citet{Hezaveh:2016} sub-halo inference has not yet been reproduced. 

In this paper, we re-analyse the ALMA observations of SDP\,81 to test the robustness of previously reported low-mass haloes. We begin by giving an overview of the ALMA observations and the calibration procedure in Section~\ref{section:data}. In Section~\ref{section:modelling}, we explain the lens model parameterisation and methodology. In Section~\ref{section:results}, we compare our lens models and perform several tests of our methodology before discussing the implications of our findings in Sections~\ref{section:discussion} and \ref{section:conclusions}. Throughout, we assume a \cite{PlanckCollaboration:2020} flat $\Lambda$CDM cosmology.\footnote{While previous work by \cite{Hezaveh:2016} and \cite{Inoue:2016} used different cosmologies, the different parameters do not change the angular diameter distances by more than 1\,\%.} 

\section{Data}
\label{section:data}

We obtained band 6 and 7 calibrated visibilities of SDP\,81 from the ALMA archive taken as part of the 2014 ALMA Long Baseline Campaign \citep{ALMAPartnership:2015}. The pipeline calibration and subsequent reduction of the band 7 data were performed using Common Astronomy Software Applications (CASA; \citealt{McMullin:2007}) and described by \cite{Stacey:2024}. The band 6 data was reduced with the same procedure. 

We did no further time or spectral averaging, resulting in data sets of 16 spectral channels of 0.5\,GHz\footnote{Four spectral windows of four channels were manipulated into 16 spectral windows of 1 channel so the data could be converted to UVFITS format. We refer to these as 16 spectral channels for simplicity.} and 2\,s integration per visibility for the band 7 data, and 8 channels of 1\,GHz and 6\,s integration for the band 6 data. This channel bandwidth results in $\leq1\%$ and $\leq3\%$ amplitude loss, respectively, at 3\,arcsec from the phase centre, adopting the formula from \cite{Bridle:1989} for a square bandpass. 2 and 6\,s time integrations both result in $\ll1$\% amplitude loss at 3\,arcsec \citep{Bridle:1989}. We set the absolute data weights and noise standard deviations by subtracting subsequent visibilities in a time series per baseline and computing the average in 1200\,s chunks for each spectral channel. 

\begin{figure*}
\centering
    \includegraphics[width=0.78\textwidth]{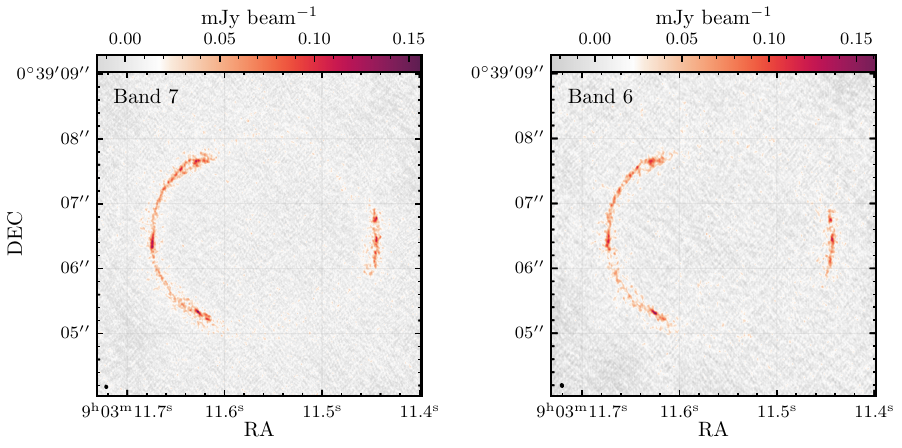}
    \vspace{-0.1cm}
    \caption{ALMA images of SDP\,81 with natural weighting of the visibilities created via forward-modelling. The synthesised beam FWHM with natural weighting is $26\times33$\,mas and $33\times42$\,mas for bands 7 (left) and 6 (right), respectively. }
    \label{fig:clean_image}
\end{figure*}

For reference, `Clean' images of the data are shown in Fig.~\ref{fig:clean_image}, which were made by convolving the maximum a-posteriori sky models (detailed in the subsequent section) with a Gaussian fit to the central lobe of the dirty beam (i.e. the synthesised beam) and adding it to an image of the residual visibilities. This forward-modelling approach results in fewer artefacts in the imaging than deconvolution with the commonly adopted CLEAN algorithm (see also \citealt{Stacey:2024b}). 

All the data is used in the subsequent analysis, although the data from the two bands are treated separately. For comparison, \cite{Hezaveh:2016} binned the data in 12\,m bins and only used visibilities with lower measured instrumental noise levels, amounting to 40\,\% of the total data. No additional time averaging was done, and the minimum and maximum $uv$ distance of the data are the same (Y. Hezaveh, private communication). \cite{Inoue:2016} performed their analysis on images produced with the CLEAN algorithm after tapering the data to an effective synthesised beam size of 0.17~arcsec. 

\section{Lens modelling}
\label{section:modelling} 


\begin{table*}
    \caption{Summary explanation of the various models tested on the band 7 data.}
    \centering
    \renewcommand*{\arraystretch}{1.12}
    \adjustbox{width=\textwidth}{
    \begin{tabular}{l | l} \hline
        Model & Description \\ \hline 
        PL & Elliptical power-law and external shear \\
        PL+MP & PL and multipoles ($m=3,4$) \\
        PL+MP+PJ & PL+MP and a pseudo-Jaffe sub-halo \\
        PL+MP+PJ\,(p) & PL+MP and a pseudo-Jaffe sub-halo fixed in position of H16 sub-halo \\
        PL+MP+PJ\,(p,m) & PL+MP and a pseudo-Jaffe sub-halo fixed in position and mass of H16 sub-halo (with H16 parameterisation) \\
        PL+MP$^{\dagger}$ & PL and multipoles ($m=3,4$) with mean parameters of H16 \\ \hline
    \end{tabular}
    }
    \label{table:model_exp}
\end{table*}

\subsection{Model parameterisation}

We employed \textsc{pronto}, a Bayesian pixellated lens modelling technique appropriate for interferometric data introduced by \citeauthor{Powell:2021} (\citeyear{Powell:2021,Powell:2022}, see also \citealt{Vegetti:2009,Rybak:2015a,Rybak:2015b,Rizzo:2018,Ritondale:2019,Ndiritu:2025}). An analysis of the smooth model for these data is detailed by \cite{Stacey:2024} and extended here to include additional lens model components. Table~\ref{table:model_exp} gives a summary explanation of these model components. The Bayesian inference methodology is detailed in \cite{Stacey:2024}. 

\subsubsection{Smooth model}
\label{section:smooth_models}

As described by \cite{Stacey:2024}, the global mass distribution of the lens is parameterised by a singular power-law ellipsoidal density profile with external shear. The convergence of the power-law density profile is described
by 
\begin{equation}
    \kappa_l(R) = \frac{3-\gamma}{2} \left ( \frac{\kappa_{0,l}}{R} \right )^{\gamma -1}\,,
    \label{eq:PL}
\end{equation} 
where $R$ is the elliptical radius ($R^{2}=q^{2}x^{2} + y^{2}$), $\gamma$ is the 3D logarithmic density slope, $q$ is the axis ratio and $\kappa_{0,l}=\sqrt{q}\ \theta_{E}$, where $\theta_E$ is the Einstein radius. The external shear is described by a strength, $\Gamma$, and position angle, $\theta_\Gamma$. The elliptical power-law plus external shear component is denoted `PL'. 

Following \cite{Stacey:2024}, we allowed for multipole perturbations that add angular variations to the PL model, which are described by
\begin{equation}
    \kappa_m(R_0,\theta) = \kappa_{0,m} \left (\frac{R_{0}}{1\ {\rm arcsec}}\right )^{-(\gamma-1)}\left[A_m \sin(m\theta)+B_m\cos(m\theta)\right]\ ,
    \label{eq:MP}
\end{equation} 
in polar co-ordinates for multipole order $m$, where $\gamma$ is the slope of the density profile in Eq.~\ref{eq:PL} and $R_0$ is the (spherical) radius ($R_0^2 = x^2 + y^2$) in terms of a scale radius of 1\,arcsec and $\kappa_{0,m}$ is the normalisation. $A_{m}$ and $B_{m}$ are dimensionless free parameters. We consider $m=3,4$ and denote this multipole model component `MP'.

\subsubsection{Sub-halo model}
\label{section:subhalo_models}

We model a low-mass halo as a sub-halo of the primary lensing galaxy (i.e. at the same redshift as the lens), considering a spherical pseudo-Jaffe (denoted as `PJ'; \citealt{Dalal:2002}) profile. We used this model for two reasons: firstly, we wish to compare our results to the previous analysis of these data by \cite{Hezaveh:2016} who used a PJ model; secondly, because a combination of tidal stripping and feedback is predicted to result in a mass density profile close to isothermal \citep{Heinze:2024}. 

The PJ component is a truncated isothermal density profile. We follow the definition of \cite{Despali:2018} whereby the 2D truncation radius ($R_t$) is a function of the sub-halo mass ($M_{s}$) and its distance from the main halo (i.e. the primary lens). The convergence is defined as 
\begin{equation}
    \kappa_{\rm s}(R_0) = \kappa_{0,s}\ R_t [ R_0^{-1} - (R_0^2 + R_t^2)^{-1/2} ]\ .
\end{equation} 
The mass is defined as
\begin{equation}
    M_{s} = 2\pi\Sigma_{c} R_t^2 \kappa_{0,s}\ ,
    \label{eq:MPJ}
\end{equation}
where $\Sigma_{c}$ is the critical surface mass density at the redshift of the lens and $\kappa_{0,s}$ is the sub-halo normalisation. The truncation radius is described by 
\begin{equation}
    R_t = R_0 \sqrt{\frac{\pi \kappa_{\rm 0,s}}{2 \xi \kappa_{\rm 0,l}}}\ ,
\end{equation}
where $\kappa_{0,l}$ is the normalisation of the PL component and $\xi$ is the impact parameter (here assumed to be 3). 

\begin{figure*}
    \includegraphics[width=0.96\textwidth]{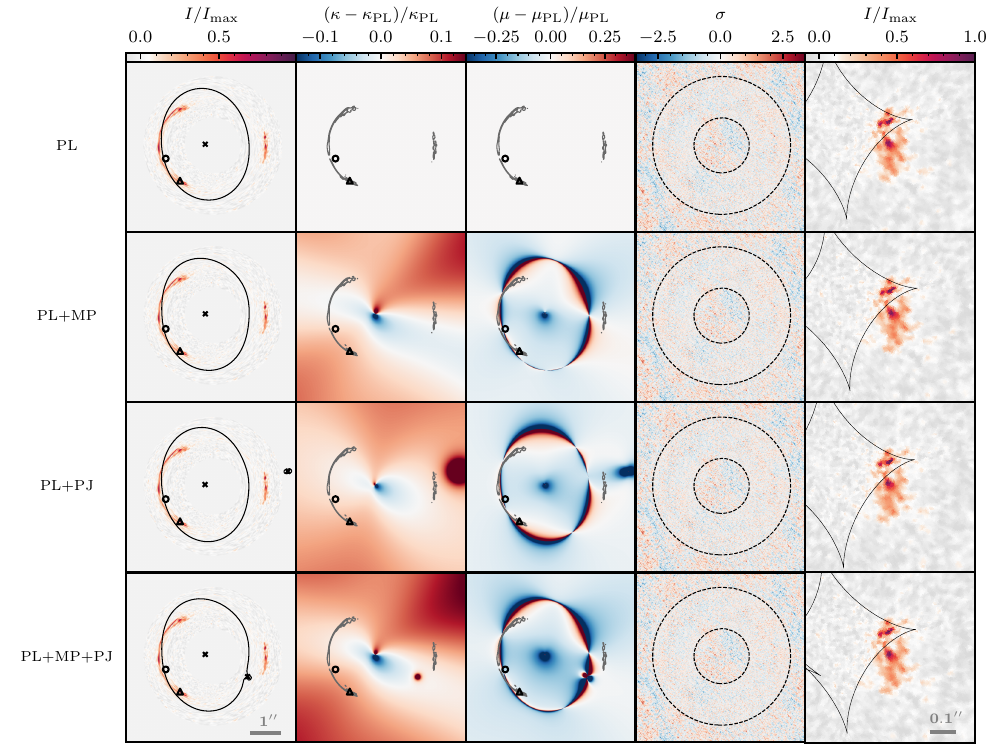}
    \caption{ Maximum {\it a posteriori} lens models for the band 7 data. First column: lens model; crosses show lens and sub-halo positions; critical curves in black. Second column: difference in convergence relative to the PL model; contours of the sky model in grey. Third column: difference in magnification relative to the PL model; contours of the sky model in black. Fourth column: image of the residuals (data$-$model) normalised by the RMS noise ($\sigma$), with the mask shown by the dashed outline. Last column: reconstructed source with Gouraud interpolation and caustics in black. The locations of I16 and H16 purported sub-haloes are shown with a diamond and a circle, respectively. }
    \label{fig:reconst}
\end{figure*}

\begin{table*}
    \caption{ Lens model parameters for the band 7 data, their $1\sigma$ uncertainties and priors. All positions are given relative to the observation phase centre. The log Bayes factor ($\ln\mathcal{K}$) is relative to the model with the highest evidence. $M_{\rm 10\,kpc}$ is not a free parameter, but is computed from the \textsc{MultiNest} chains. The $M_{s}$ definition is described in Eq.~\ref{eq:MPJ}. Positions are given relative to the phase centre (J2000 09:03:11.61 +00:39:06.70) and angles are defined east of north. Parameter definitions follow the Code-independent Organised Lens Standard \citep{Galan:2023}. An extended version of this table is presented in Table~\ref{table:evidence_ext} of the Appendix.}
    \setlength{\tabcolsep}{2pt}
    \renewcommand*{\arraystretch}{1.12}
    \begin{tabularx}{\textwidth}{ C{3.0cm} C{2.9cm} C{2.9cm} C{2.9cm} C{2.9cm} C{2.9cm}} \hline 
                                         & PL            & PL+MP         & PL+PJ         & PL+MP+PJ      & Prior    \\  \hline \noalign{\smallskip}
    $x_l$ ($''$)                         & $0.542\pm0.001$ & $0.560\pm0.001$ & $0.548\pm0.001$ & $0.560\pm0.001$ & $\mathcal{U}(0.5,0.6)$ \\ 
    $y_l$ ($''$)                         & $-0.170\pm0.003$ & $-0.149\pm0.002$ & $-0.163\pm0.001$ & $-0.149\pm0.002$ & $\mathcal{U}(-0.2,-0.1)$ \\ 
    $\theta_E$ ($''$)                    & $1.609\pm0.004$ & $1.611\pm0.001$ & $1.613\pm0.001$ & $1.611\pm0.001$ & $\mathcal{U}(1.4,1.8)$\\ 
    $\log\,(M_{\rm 10\,kpc}/{\rm M_{\odot}})$       & $11.636\pm0.004$ & $11.639\pm0.004$ & $11.637\pm0.003$ & $11.640\pm0.004$ & - \\
    $q$                                  & $0.794\pm0.005$ & $0.832\pm0.003$ & $0.805\pm0.004$ & $0.832\pm0.003$ & $\mathcal{U}(0.7,0.9)$ \\ 
    $\theta_q$ ($^\circ$)                & $13\pm2$ & $6\pm1$ & $9.3\pm0.5$& $6\pm1$ & $\mathcal{U}(-10,30)$ \\ 
    $\gamma$                             & $1.97\pm0.01$ & $2.00\pm0.01$ & $2.00\pm0.01$& $2.01\pm0.02$ & $\mathcal{U}(1.9,2.1)$ \\ 
    $\Gamma$                             & $0.030\pm0.001$ & $0.037\pm0.002$ & $0.038\pm0.002$ & $0.037\pm0.002$ & $\log\mathcal{U}(0.001,0.1)$ \\  
    $\theta_\Gamma$ ($^\circ$)           & $-9\pm2$ & $9\pm1$ & $3\pm1$& $9\pm1$ & $\mathcal{U}(-20,20)$ \\ 
    $A_3$                                & -             & $0.0018\pm0.0004$ & -             & $0.0018\pm0.0004$ & $\mathcal{G}(0.00,0.01)$ \\ 
    $B_3$                                & -             & $0.0034\pm0.0003$ & -             & $0.0034\pm0.0003$ & $\mathcal{G}(0.00,0.01)$ \\ 
    $A_4$                                & -             & $-0.0032\pm0.0006$ & -            & $-0.0032\pm0.0006$ & $\mathcal{G}(0.00,0.01)$ \\ 
    $B_4$                                & -             & $0.0041\pm0.0007$ & -             & $0.0041\pm0.0006$ & $\mathcal{G}(0.00,0.01)$ \\ 
    $x_{\rm s}$ ($''$)                 & -             & -                & $3.1\pm0.1$& $0.9\pm1.5$ & $\mathcal{U}(-2.0,3.5)$ \\ 
    $y_{\rm s}$ ($''$)                 & -             & -                & $0.26\pm0.02$ & $-0.4\pm1.5$ & $\mathcal{U}(-3.0,2.5)$ \\ 
    $\log\,(M_{s}/{\rm M_{\odot}})$ & -         & -             & $9.6\pm0.1$ & $<8.1$ & $\log\mathcal{U}(6,10.4)$ \\ \noalign{\smallskip}\hline
    $\ln\mathcal{K}$                        & $-28$     & $0$        & $-6$     & $0$        \\  \hline \noalign{\smallskip}
    \end{tabularx} 
    \label{table:evidence}
\end{table*}

\section{Analysis of real data}
\label{section:results}

The goal of this work is to compare our lens modelling analysis with
previous studies by \cite{Hezaveh:2016} and \cite{Inoue:2016} (H16 and I16, hereafter). First, we report the results of our smooth modelling, including flexibility for angular and radial variations in the mass distribution (Section~\ref{section:smooth}). Second, we test the reported detections of low-mass haloes in these data by including sub-halo model components (Section~\ref{section:search}). We then test the effects of regular and adaptive source gridding (Section~\ref{section:gridding}).

The summary statistics for the posterior distributions of the lens model parameters from \textsc{MultiNest} are given in Table~\ref{table:evidence}, and the maximum {\it a posteriori} models are shown in Fig.~\ref{fig:reconst}. Fig.~\ref{fig:cornerplot} in the Appendix is a corner plot showing the two-dimensional posterior probability distributions. Heat maps comparing the log Bayes factors of the various models tested are shown in Figs.~\ref{fig:heatmap}--\ref{fig:heatmap_H16_regulargrid}. Extended tables with summary statistics for the various models tested are in the Appendix.

As detailed in \citealt{Stacey:2024}, to prevent overfitting, it is necessary to impose regularisation when solving for the source surface brightness. We considered three forms of regularisation: curvature, gradient and area-weighted gradient methods \citep{Suyu:2006,Stacey:2024}. Additionally, we considered different numbers of pixels cast into the source plane to impose another form of regularisation \citep{Bayer:2023}. We performed Multinest sampling of the PL model for the band 7 data, with 1, 2, 3 and 4 cast pixels, for each of the different regularisation types. We found that 3 pixels and gradient regularisation resulted in the highest Bayesian evidence, and this is what we used for all the subsequent analysis.

\subsection{Smooth model}
\label{section:smooth}

\subsubsection{Power-law model}

Our inferred PL model parameters are similar to the PL models inferred from previous studies of this lens using the visibility data from baselines $<2$\,km \citep{Rybak:2015a} or `Clean' images of the data at $\approx30$\,mas resolution \citep{Dye:2015,Inoue:2016,Tamura:2015,Wong:2015}. We find a power-law slope, ellipticity, ellipticity position angle, shear, and shear position angles consistent within $1\sigma$ with \citeauthor{Dye:2015} who used a pixellated modelling of the Briggs-weighted Clean band 6 and band 7 images. We also find ellipticity, ellipticity position angle, shear, and shear position angle consistent within $1\sigma$ of those inferred by \citeauthor{Inoue:2016} for a singular isothermal ellipsoid and external shear lens model, and using point-sources extracted from a Clean image as constraints.

H16 reported the summary statistics only for PL+MP+PJ, not for PL and PL+MP individually, so a fair comparison with our PL model cannot be made. We compare our PL+MP component parameters in the following section.

\subsubsection{Multipoles}
\label{section:multipoles}

PL+MP was preferred with a log Bayes factor ($\ln\mathcal{K}$) of 28 relative to PL (i.e. $\ln\mathcal{K} = \ln\mathcal{E}_{\rm PL+MP} - \ln\mathcal{E}_{\rm PL}$, where $\mathcal{E}$ is the Bayesian model evidence), suggesting angular complexity in the lens model beyond the PL model. The parameter posteriors and summary statistics differ between PL and the PL component of PL+MP (see Fig~\ref{fig:cornerplot} and Table~\ref{table:evidence}). Several PL model parameters shift by $\geq1\sigma$ with the inclusion of MP, most significantly the lens centre and position angles of the external shear (by $>3\sigma$). A comparison of PL and PL+MP for this lens was previously reported by \cite{Stacey:2024} for these band 7 data with the same methodology used here. We refer to \cite{Stacey:2024} for a more detailed analysis of the multipole structure. 

Besides this work and that of \cite{Stacey:2024}, the only other study in the literature to include an MP component in their lens model was H16. Despite shifts in PL component parameters with the inclusion of multipoles, we find significant disagreement with the parameters found by H16. We compute the mass enclosed within 10\,kpc for our inferred lens model posterior and find $\log (M_{\rm 10kpc}/{\rm M_{\odot}}) = 11.64$ (see Table~\ref{table:evidence}), which is consistent within 3$\sigma$ of the enclosed mass reported by H16, accounting for differences in cosmology. However, after accounting for differences in parameter definitions, we find that the PL model we infer has significantly different ($>10\sigma$) external shear strength and ellipticity, as well as a significantly different ellipticity position angle. The external shear of the H16 PL is an order of magnitude smaller, the ellipticity is significantly higher ($>10\sigma$), and the ellipticity position angle is 70\,deg different. Furthermore, H16 inferred significantly higher $m=3$ multipole coefficient amplitudes than we find, albeit with much larger uncertainties. After accounting for different multipole parameterisations (Y. Hezaveh, private communication), the H16 mean multipole parameters are $A_3=0.003\pm0.003$; $B_3=0.013\pm0.003$; $A_4=0.006\pm0.005$; $B_4=0.003\pm0.006$ for our parameterisation (Eq.~\ref{eq:MP}). We find $m=3$ multipole amplitudes ($\eta_m=\sqrt{A_m^2+B_m^2}$), a factor of $3.3\pm0.8$ times smaller than H16. The $m=4$ amplitudes and position angles ($\phi_m=(1/m)\arctan{B_m/A_m}$) are consistent within their $1\sigma$ uncertainties. 

To confirm that we did not converge to a local minimum in the posterior parameter space, we re-ran our initial PL optimisation beginning from the lens parameters inferred by H16. The optimisation ultimately converged to the same maximum {\it a posteriori} parameters that we initially inferred. Additionally, we re-ran the \textsc{MultiNest} sampling with all uniform priors large enough to encompass the mean and uncertainties of the MP and PL parameters inferred by H16, but it did not change the inferred evidence or summary statistics.

H16 reported that a PL model similar to ours and others reported in the literature \citep{Dye:2015,Inoue:2016,Tamura:2015,Wong:2015} gave a comparably good fit to the data and did not significantly alter their sub-halo inference. Therefore, as an additional test, we fixed the lens model to our maximum {\it a posteriori} PL and the mean MP parameters reported by H16 (denoted PL+MP$^{\dagger}$) and optimised only the regularisation hyperparameter to allow the source to focus. This allowed us to test these multipoles directly and how they might impact the sub-halo inference. This model resulted in larger image-plane residuals ($>5\sigma$) compared to our best-fit PL+MP ($<3\sigma$, see Fig.~\ref{fig:reconst}). We then considered whether a different PL component could produce a better model in combination with MP$^{\dagger}$ by allowing the PL component to optimise. Still, this resulted in $>5\sigma$ features in the image-plane residuals. To quantify this goodness of fit, we ran \textsc{MultiNest} for PL+MP$^{\dagger}$ and our best fit PL+MP model with all model parameters fixed except for the source regularisation hyperparameter so that we can compare the model evidence. We inferred a log Bayes factor of -697 for PL+MP$^{\dagger}$ relative to our best PL+MP, confirming that the H16 MP parameters are a poorer fit to the data. 

\subsection{Sub-halo search}
\label{section:search}

\subsubsection{Sub-halo search with our smooth model}

To determine whether the data favour a sub-halo, we included a parametric sub-halo component in our models (PJ, as detailed in Section~\ref{section:subhalo_models}). We explored the posterior parameter space with \textsc{MultiNest}, allowing the PJ to be anywhere on the lens plane with a log-prior on the mass of $\log (M_{\rm sub}/{\rm M_{\odot}}) \in [6,10.4]$. The smooth model parameters were free with the same priors as for the smooth model inference. We found that the PL+PJ is preferred by a log Bayes factor of 22 relative to PL. While the PJ component is constrained in mass and location, it is in a different location than reported in the previous studies and coincides with a positive convergence feature in the PL+MP model (Fig.~\ref{fig:reconst}). The parameter summary statistics for PL+PJ show that the PL component parameters move closer to the PL component parameters of PL+MP, suggesting that the PL and PJ components conspire to induce a similar, non-localised lensing effect. This is also indicated by the apparent convergence and magnification differences in Fig.~\ref{fig:reconst}.

Previous work has found that incorrect conclusions about the presence of sub-halos may be inferred if angular structure isn’t included in the smooth model \citep{O'Riordan:2024}. To avoid this, we include a multipole component in subsequent sub-halo searches to account for angular as well as localised mass structure. For posterior sampling with \textsc{MultiNest}, the smooth model and the PJ component parameters were free with the same priors as before. Comparing the two models, PL+MP and PL+MP+PJ, we found no difference in Bayesian evidence (see Table~\ref{table:evidence}), meaning the data does not favour one model over the other. The PL and MP component parameters were consistent across both models, with differences falling within $1\sigma$ uncertainties (see Fig.~\ref{fig:cornerplot} and Table~\ref{table:evidence}). The PJ component position was not constrained (see Fig.~\ref{fig:cornerplot}), and we infer an upper mass limit (at the 86th percentile) of $10^{8.1}$\msol. Therefore, we infer no evidence in favour of a sub-halo when multipoles are included in the model.

This is in contradiction to the inference of the $\approx10^{9}~{\rm M_{\odot}}$ sub-halo by H16. Therefore, to more directly test the sub-halo reported by H16, and to ensure we sufficiently explore the relevant parameter space, we fixed the sub-halo’s position to the coordinates given by H16 (this model is denoted PL+MP+PJ(p)). The sub-halo mass was left as a free parameter, with the same prior bounds as before. The PL+MP parameters and regularisation remained free with the same priors as before. In this case, we found that the PJ mass tended to the lower edge of the prior and resulted in no improvement in the Bayesian evidence over PL+MP. We infer an 86th percentile upper limit of $10^{7.8}$\msol, indicating that the data do not favour a sub-halo in this location. 

To better quantify the level of exclusion of the H16 sub-halo, we directly fixed both the sub-halo mass and position to that found by H16 while leaving the PL+MP parameters free to allow them to compensate for the sub-halo component. This model is denoted PL+MP+PJ(p,m). 
We note that the PJ parameterisation implemented by H16 is slightly different from our analysis: the truncation radius used by H16 is defined relative to the ratio of velocity dispersions of isothermal spheres of the masses of the primary lens and PJ. This results in a deflection equivalent to a less massive sub-halo in the \cite{Despali:2018} definition. To ensure a fair comparison, we implemented the exact parameterisation used by H16 (Y. Hezaveh, private communication). With \textsc{MultiNest} posterior sampling, we found that the PL+MP+PJ(p,m) model is disfavoured relative to PL+MP, with a log Bayes factor of 13. Still, we do not find agreement with the PL or MP component parameters of H16. A heat map showing the log Bayes factors of the models tested on the band 7 data is shown in Fig.~\ref{fig:heatmap}. Summary statistics for PL+MP+PJ(p) and PL+MP+PJ(p,m) lens models are reported in Table~\ref{table:evidence_ext} in the Appendix.

While the band 7 data has a better signal-to-noise ratio and angular resolution, we note that H16 found improved evidence for a sub-halo by combining the band 6 and 7 data sets. We explicitly choose not to combine these data sets due to the different flux density, source structure (see \citealt{Rybak:2015a}) and possible astrometric offsets. Still, comparing our findings for the two data sets helps test the sensitivity of the sub-halo detection to phase and amplitude errors in the data. Therefore, we also searched for evidence of a sub-halo in the band 6 data, but we found no improvement in the Bayesian evidence of PL+MP+PJ or PL+MP+PJ(p) relative to PL+MP (i.e. $\ln\mathcal{K}=0$). We infer 86th percentile upper limits of $10^{8.0}$\msol for the mass of a sub-halo in the location inferred by H16, indicating that neither band 6 or 7 data sets favour the H16 sub-halo. 

\subsubsection{Sub-halo search with the H16 smooth model}

As a final test, we investigated whether the different multipole coefficients used by H16 could impact the sub-halo inference. Specifically, we considered whether the combination of the H16 smooth model parameters and sub-halo might conspire to improve the overall model fit. Using the band 7 data, we fixed the smooth model components to our maximum {\it a posteriori} PL and multipoles reported by H16 (PL+MP$^{\dagger}$), leaving only the regularisation free to allow the source to focus (as described in Section~\ref{section:multipoles}). We added a PJ component with mass prior bounds of $\log (M_{\rm sub}/{\rm M_{\odot}}) \in [6,10.4]$ and position prior bounds of the image plane. This allowed us to test whether a sub-halo with similar properties to that found by H16 could be recovered under these assumptions.

With this setup, PL+MP$^{\dagger}$+PJ is strongly favoured relative to PL+MP$^{\dagger}$ with a log Bayes factor of 365. This is not unsurprising, because we established that PL+MP$^{\dagger}$ produces $>5\sigma$ residuals, so it is expected that the additional degrees of freedom will act to reduce these residuals. However, the inferred mass and position of the PJ component are $\log\,M_{\rm sub} = 9.93\pm0.03$, $x_s = -1.25\pm0.04$, $y_s=-1.64\pm0.04$. These differ from the H16 sub-halo by more than $10\sigma$, indicating that it is not the same model inferred by H16. Therefore, we also tested the H16 sub-halo directly by fixing both its mass and position. This model, PL+MP$^{\dagger}$+PJ(p,m), was preferred by a log Bayes factor of 75 relative to PL+MP$^{\dagger}$. This is equivalent to $12\sigma$, higher than the $6\sigma$ inferred by H16. 

\begin{figure}
    \begin{subfigure}[b]{0.5\textwidth}
        \includegraphics[width=\linewidth]{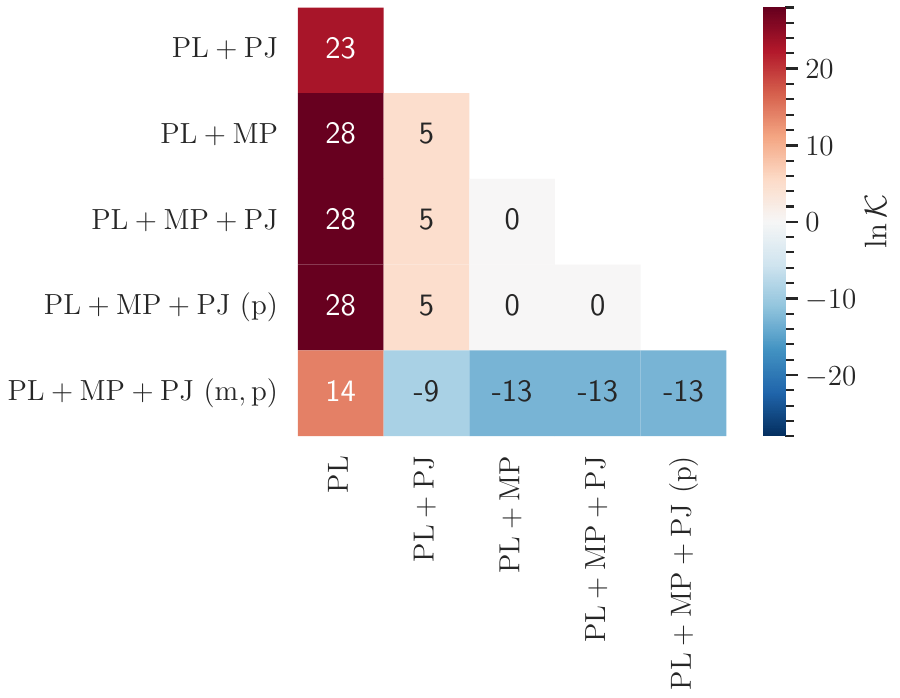}
        \caption{Models tested on the real band 7 data.}
        \label{fig:heatmap}
    \end{subfigure}

    \begin{subfigure}[b]{0.5\textwidth}
        \vspace{10pt}
        \centering
        \includegraphics[width=0.85\linewidth]{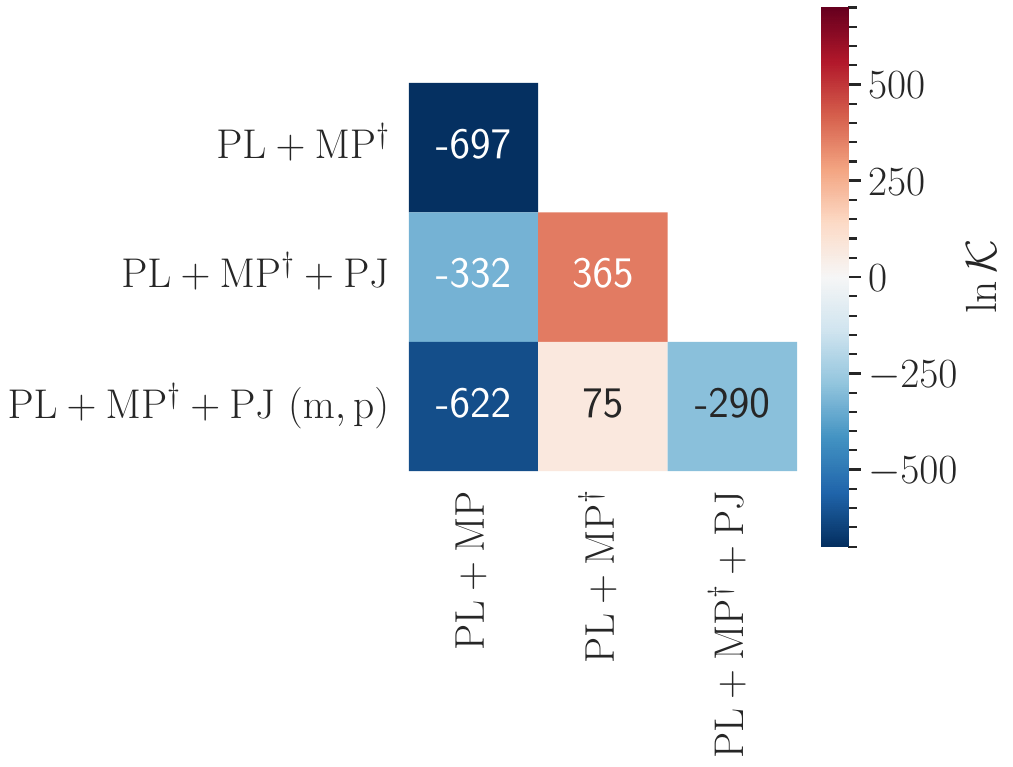}
        \caption{Models tested on the real data with H16 multipole coefficients (MP$^{\dagger}$). The smooth models are fixed, and the regularisation is the only free parameter.}
        \label{fig:heatmap_H16}
    \end{subfigure}
    \caption{Heat map showing the log Bayes factors of the models tested on the real band 7 data (y-axis model relative to x-axis model). }
\end{figure}

Nevertheless, as shown in Fig.~\ref{fig:heatmap_H16}, all these models based on the H16 smooth lens model (PL+MP$^{\dagger}$) are very strongly disfavoured relative to our maximum {\it a posteriori} PL+MP model, with log Bayes factors in the range $-332$ to $-697$. Therefore, these results still indicate no statistical preference for a sub-halo. The relative increase in evidence for the H16 sub-halo in combination with the H16 MP component indicates that the MP component may have led to a spurious sub-halo inference.

\subsection{Regular vs adaptive source gridding}
\label{section:gridding}

A key methodological difference between this work and H16 lies in the choice of the source reconstruction grid.  In our analysis, we use a magnification-adaptive source grid, as described in \cite{Vegetti:2009}, whereas H16 employed a regular grid with an arbitrary 10\,mas pixel scale. Each approach has its advantages and limitations. Adaptive grids provide greater flexibility in highly magnified regions, potentially leading to overfitting if not properly regularised. Regular grids, on the other hand, may either overfit or underfit depending on the selected pixel scale, which is not dynamically adjusted to the lensing geometry. As a result, the sensitivity to sub-halo perturbations may differ between the two methods: an adaptive grid might smooth over small perturbations that a regular grid could capture, or vice versa, depending on the local structure of the lensing potential and the chosen source grid resolution.

To assess whether the choice of source grid influences sub-halo detection, we repeated our analysis using a regular source grid with a fixed 10~mas pixel scale, matching the setup used by H16. This allowed us to test whether similar evidence for the H16 sub-halo (or any sub-halo) could be recovered in this scenario.

Using the regular grid, we found that the PL+MP+PJ model was disfavoured relative to PL+MP, with a log Bayes factor of 5. In this case, the PJ component was unconstrained in both mass and position. We also tested a model with the PJ fixed at the position of the H16 sub-halo. This configuration was similarly disfavoured, with a log Bayes factor of 6, and yielded an 86th percentile upper limit on the PJ mass of $10^{7.6}$\msol. The model where both the mass and position of the PJ were fixed to the H16 values (PL+MP+PJ(p,m)) was more strongly disfavoured than for the adaptive grid, with a log Bayes factor of 70 relative to PL+MP. 

\begin{figure}
    \begin{subfigure}[b]{0.5\textwidth}
        \centering
        \includegraphics[width=0.9\linewidth]{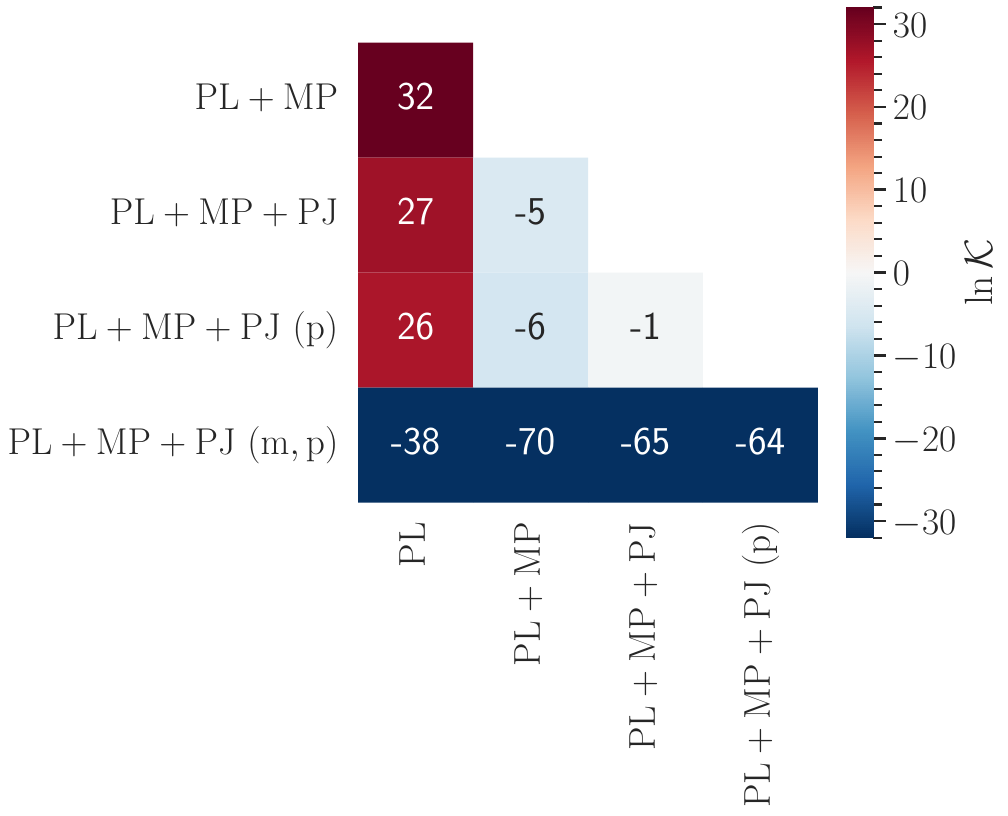}
        \caption{Same as Fig.~\ref{fig:heatmap}, using a regular source grid.}
        \label{fig:heatmap_regulargrid}
    \end{subfigure}

    \begin{subfigure}[b]{0.5\textwidth}
        \vspace{10pt}
        \centering
        \includegraphics[width=0.85\linewidth]{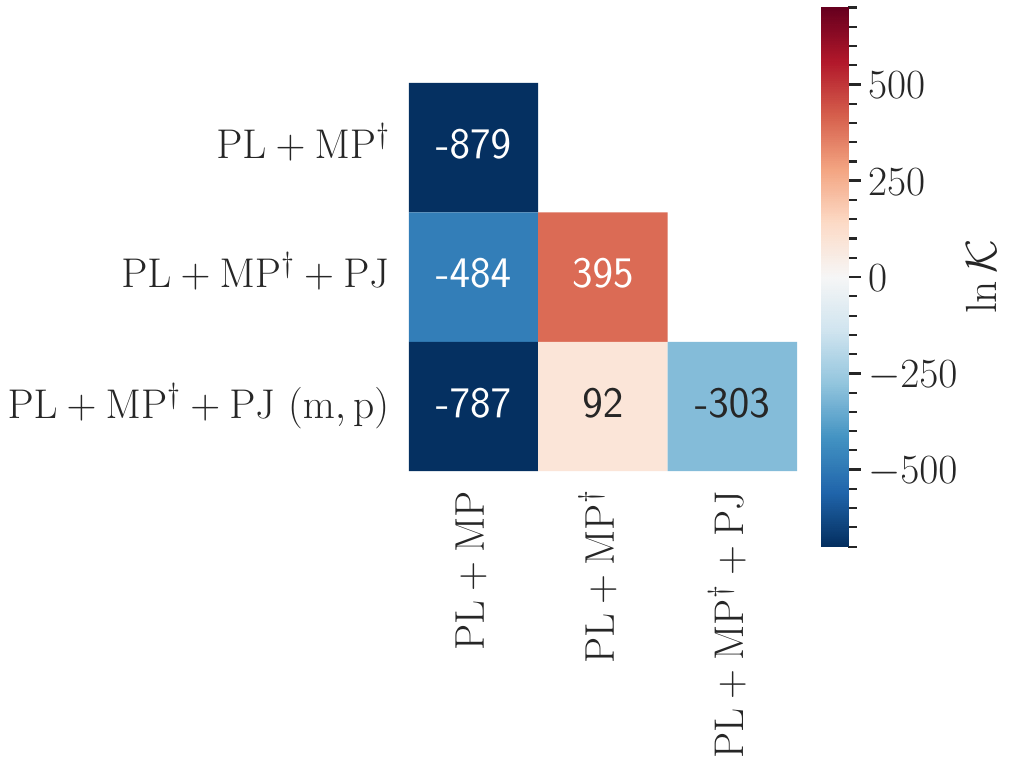}
        \caption{Same as Fig.~\ref{fig:heatmap_H16}, using a regular source grid.}
        \label{fig:heatmap_H16_regulargrid}
    \end{subfigure}

\caption{Heat map showing the log Bayes factors of the models
tested on the real band 7 data (y-axis model relative to x-axis
model), using a regular source grid of 10~mas.}
\end{figure}

The larger overall Bayes factors found using the regular grid (compared to the adaptive grid) likely reflect the stronger effective regularisation imposed by the fixed 10~mas resolution, in contrast to the more flexible Delaunay-based adaptive grid. For visual comparison, Fig.~\ref{fig:heatmap_regulargrid} shows a heat map of the log Bayes factors across all tested models for a regular grid.

As a further investigation, we performed these same tests with the smooth model inferred by H16 (PL+MP$^{\dagger}$), to exclude the possibility that this model succeeds for a regular grid but not an adaptive grid. We fixed the smooth model parameters, with the only free parameter being the regularisation to allow the source to focus (i.e. the same test telegraphed by Fig~\ref{fig:heatmap_H16}). As shown in Fig.~\ref{fig:heatmap_H16_regulargrid}, these models are also all very strongly disfavoured relative to our best smooth model for a regular source grid ($\ln\mathcal{K}=-484$ to $-879$). 

As for the adaptive grid, the evidence increases with the inclusion of the H16 sub-halo relative to the H16 smooth model (by a factor of 92). While the Bayes factors differ, the overall result is the same as for the adaptive grid. Therefore, we do not find evidence that the H16 sub-halo was an artefact produced by the regular source grid, nor do we find evidence that the difference in source grid type could change our interpretation.

\section{Analysis of mock data}

We tested a number of our findings with realistic mock data. One of our findings was angular structure, and an apparent degeneracy between that and a single massive sub-halo, which we explore in Sections~\ref{section:smooth_mock} and~\ref{section:subhalo_mock}. Additionally, we do not find agreement with the previous study by H16 with a similar approach, so we tested whether we could detect such a sub-halo (Section~\ref{section:subhalo_mock}) and test the effects of phase errors (Sections~\ref{section:phaserr}). 

We made realistic mock data by overwriting the real band 7 visibility data with a Fourier transform of the maximum {\it a posteriori} lens model and adding Gaussian noise per channel per correlation as measured in the real visibility data in 1200\,s chunks. We modelled the mock data with the same procedure as for the real band 7 data, including the same priors for Multinest sampling (Table~\ref{table:evidence}).

\subsection{Smooth model}
\label{section:smooth_mock}

With mock data generated from our maximum {\it a posteriori} PL+MP lens model, we verify that the inferred multipole amplitudes can be reliably recovered. The posterior distributions obtained from the mock data are consistent with those from the real data, within 1$\sigma$ (see Fig.~\ref{fig:mp_posteriors} and Table~\ref{table:evidence_ext_mp} in the Appendix). We also find a log Bayes factor of 27 in favour of the PL+MP model over the PL model, closely matching the value of 28 obtained from the real observations. This agreement indicates that the detected multipole features are not the result of noise artefacts but are instead genuine structures present in the data.

\subsection{Sub-halo search}
\label{section:subhalo_mock}

Firstly, we assessed whether the similar evidence for PL+MP and PL+PJ for the real data could be reproduced. We suspected that this apparent degeneracy could be because a single massive sub-halo can induce a non-localised lensing effect and mimic a multipole effect. Therefore, we tested this using the mock data with ground truth PL+MP, introduced in Section~\ref{section:smooth_mock}. From sampling with \textsc{MultiNest}, we found PL+PJ is also favoured by 24 relative to PL; a similar, but slightly lower significance than for PL+MP (Fig.~\ref{fig:heatmap_mock_MP}; summary statistics are given in Table~\ref{table:evidence_ext_mp} in the Appendix. ). This is consistent with what we found for the real data. As there is no PJ in the ground truth, this confirms that a single massive sub-halo can mimic the lensing effect of multipoles. In this case, the PJ is constrained at the north-east corner of the image plane (edge of the prior) and is very massive ($\approx10^{10}$~\msol), where it has a non-localised effect. Although these are not the same mass and location inferred for the real data, most of the PL component parameters move closer to those in the PL+MP model, as we observed for the real data.

We then considered whether our lack of evidence for the H16 sub-halo was due to methodological differences (e.g. source regularisation or posterior inference) by performing the sub-halo search, as in Section~\ref{section:search}, on mock data with ground truth PL+MP+PJ(p,m). The mock lens model was generated from the maximum {\it a posteriori} PL+MP source, lensed by the maximum {\it a posteriori} PL+MP plus a PJ component. The PJ sub-halo had the mass, location and parameterisation as reported by H16. We performed sampling with \textsc{MultiNest} of the mock data with and without the H16 sub-halo (including multipoles) to compute the models' evidence, marginalising over the source and smooth model parameters. A Heat map comparing the log Bayes factors is shown in Fig.~\ref{fig:heatmap_mock_PJ}. 

For the mock data where the ground truth is PL+MP+PJ(p,m), we found that the PL+MP+PJ model has the highest evidence relative to PL and PL+MP. PL+MP is favoured by a log Bayes factor of 13 relative to PL, while PL+MP+PJ(p,m) is favoured by a log Bayes factor of 29 relative to PL+MP. This suggests that we would find evidence in favour of the H16 PJ with a log Bayes factor of 29 relative to PL+MP, should it exist in the real data. The posterior mass and position of the sub-halo are consistent within 3$\sigma$ uncertainties of the ground truth (see Table~\ref{table:evidence_ext_pj} for summary statistics). Therefore, we conclude that our inability to find evidence in favour of the H16 sub-halo in the real data is not a consequence of different methodologies.

\begin{figure}

    \begin{subfigure}[b]{0.5\textwidth}
        \centering
        \includegraphics[width=0.55\linewidth]{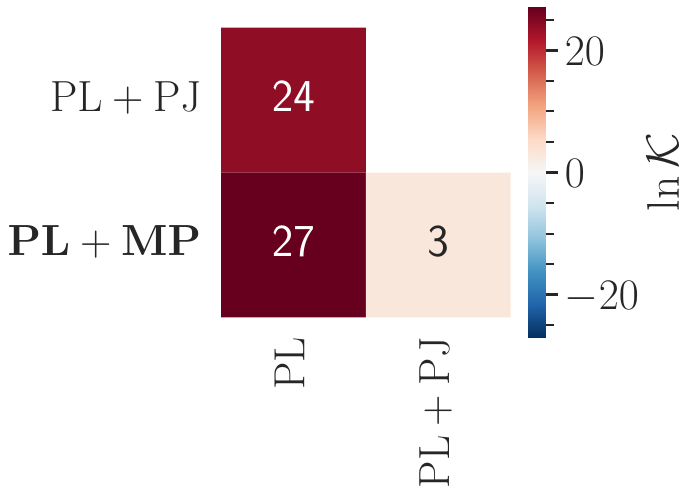}
        \caption{Models tested on mock band 7 data where the ground truth is PL+MP.}
        \label{fig:heatmap_mock_MP}
    \end{subfigure}
    \begin{subfigure}[b]{0.5\textwidth}
            \vspace{10pt}
        \centering
        \includegraphics[width=0.72\linewidth]{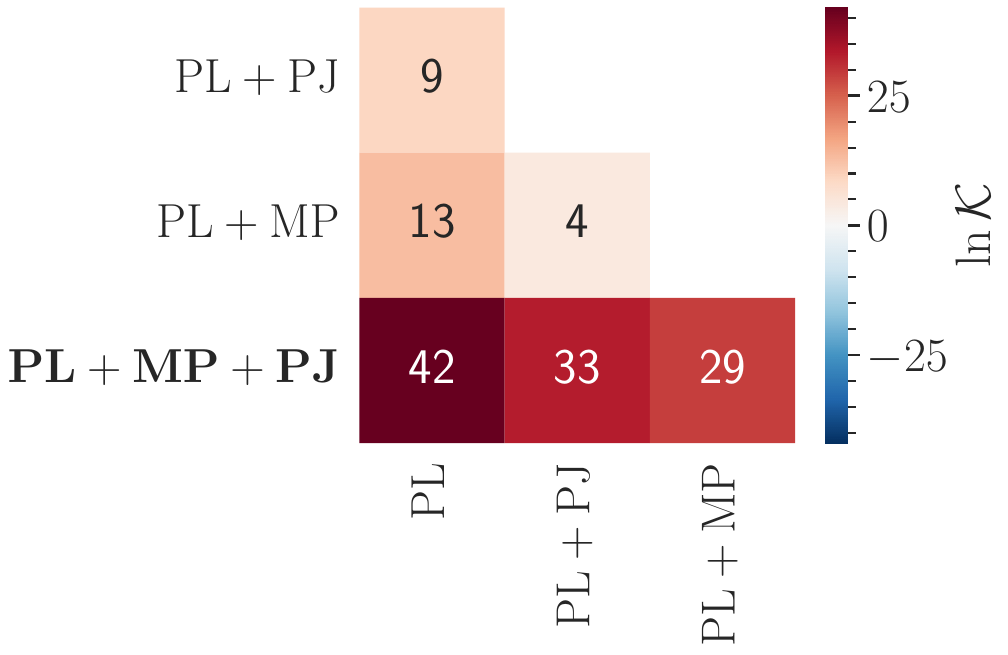}
        \caption{Models tested on mock band 7 data where the ground truth is PL+MP+PJ\,(p,m).}
        \label{fig:heatmap_mock_PJ}
    \end{subfigure}

    \caption{Heat map showing the log Bayes factors of the models tested on mock band 7 data (y-axis relative to x-axis).}
\end{figure}

\subsection{Phase errors}
\label{section:phaserr}

In the analysis thus far, we have only considered the effect of instrumental noise on the model evidence. However, antennas may have residual systematic errors as a result of transferring gain solutions from the phase calibrator to the target. The magnitude of these phase errors\footnote{We consider only phase errors here, which we expect to be more relevant than amplitude errors since phase errors redistribute the surface brightness on the sky.} depends on the specifics of the observations (e.g. distance from target to calibrator, weather conditions, flux density of the calibrator and target, baselining precision). Residual phase errors can be corrected through self-calibration, provided there is a sufficient signal-to-noise ratio per antenna (e.g., \citealt{Spingola:2018}). However, in the case of SDP\,81, the low signal-to-noise ratio per antenna means that self-calibration solutions are only possible for long time intervals (see \citealt{Stacey:2024}). Therefore, any phase errors on shorter timescales ($\lesssim1$\,hr) are unknown and uncorrected.

With this in mind, it is plausible that phase errors could induce artefacts in the data, reducing the sensitivity of the data to a sub-halo, as this method relies on the surface brightness accuracy of lensed images. Therefore, we tested whether phase errors could reduce the evidence for a sub-halo by generating mock data sets. These mock data sets had a ground truth of PL+MP+PJ(p,m) (as detailed in Section~\ref{section:subhalo_mock}) with systematic phase offsets added to each antenna, drawn from a Gaussian distribution. This assumption may not be representative of true antenna-based phase errors for ALMA that, in addition to being observation-dependent, are also time-dependent and baseline-length-dependent (see \citealt{Maud:2022,Maud:2023}). However, this is a simple test of the effect of phase errors on our mock data and is the same assumption made by H16. 

As the actual phase errors in the real data are unknown, we tested standard deviations of 5 and 10~deg. We modelled the mock data with PL+MP and PL+MP+PJ. In both cases, we recover strong evidence in favour of the sub-halo. The log Bayes factor in favour of PL+MP+PJ is 29 relative to PL+MP with 5 deg phase errors, consistent with the mock without phase errors (see Fig.~\ref{fig:heatmap_mock_PJ}), while the evidence decreases to 25 for 10 deg. We also created mock data for a standard deviation of 20~deg, but the resulting decoherence made it visibly inconsistent with the real data.

As the evidence in favour of PL+MP+PJ decreases only mildly when random phase errors are included, we conclude that these do not have a significant impact on any sub-halo inference and differences in approach to self-calibration cannot account for the lack of evidence we find for the H16 sub-halo. 

On the other hand, one could ask whether random phase errors could lead to spurious evidence in favour of a sub-halo. We tested this scenario by generating mock band 7 data with a ground truth of PL+MP (i.e. no sub-halo) and adding random phase errors, as before. We modelled these mock data with PL+MP and PL+MP+PJ and performed posterior sampling with \textsc{MultiNest}. However, the log Bayes factors for PL+MP+PJ were $<1$ relative to PL+MP for these mock data sets, indicating no preference for a sub-halo. This suggests that random phase errors would not lead to spurious inferences of sub-haloes for the real data, also indicating that the H16 inference cannot be attributed to this origin.

\section{Discussion}
\label{section:discussion}

\subsection{Sub-halo or multipoles?}
\label{section:degeneracy}

The models of the real band 7 data with the highest evidence include multipoles. PL+PJ is favoured relative to the PL model, but not relative to PL+MP+PJ. The uncertainties in the log Bayes factors reported by \textsc{MultiNest} are typically $<1$, but, as explored by \cite{Nelson:2020}, these uncertainties may be underestimated, and log Bayes factors of $\approx3$ may not be conducive to reliable model selection. The log Bayes factor of the PL+MP and PL+PJ models is 6, more than what is commonly ascribed to stochasticity.

\begin{figure}
    \includegraphics[width=0.46\textwidth]{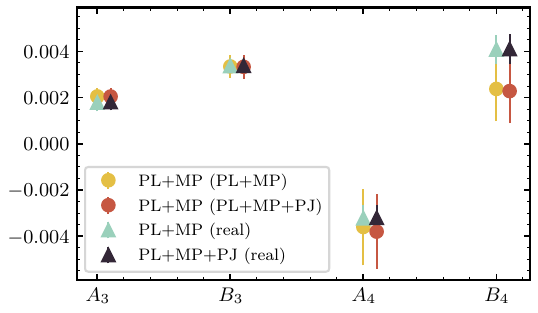}
    \caption{Multipole coefficient mean and 1$\sigma$ uncertainties for the real and mock data. Yellow and red circles show the posteriors for the mock data with ground truth PL+MP and PL+MP+PJ, respectively; cyan and black triangles show the posteriors for PL+MP and PL+MP+PJ models tested on the real data. The multipole coefficients are consistent within their 1$\sigma$ uncertainties.}
    \label{fig:mp_posteriors}
\end{figure}

However, we find that the multipole coefficient posteriors are consistent within 1$\sigma$ for models tested on the real data and mock data with ground truths PL+MP+PJ and PL+MP (Fig.~\ref{fig:mp_posteriors}), suggesting that the multipoles are largely stable to the inclusion of a sub-halo in both the model and ground truth. Additionally, with mock data with ground truth PL+MP, we found PL+PJ produced similar (but slightly lower) log Bayes factors than PL+MP, consistent with the real data, suggesting that a massive PJ at the edge of the field can induce a non-localised effect to compensate for angular structure. We note that the convergence and magnification differences of PL+MP and PL+PJ shown in Fig.~\ref{fig:reconst} have some similar features, which further hints that the PJ produces similar deflections in combination with the smooth model parameters.  
These findings favour the multipole scenario and suggest that a large sub-halo (in combination with the smooth model parameters) can induce angular perturbations in the lens potential similar to multipoles. Our results are in line with recent theoretical work that has found that the presence of angular structure induces spurious `detections' of low-mass haloes if not accounted for in the lens model \citep{O'Riordan:2024} and can fully account for the flux ratio anomalies in lensed quasar images commonly attributed to a population of low-mass haloes \citep{Cohen:2024}. 

\subsection{Differences with previous work}

We do not find evidence in favour of the sub-haloes reported by I16 or H16 in our re-analysis of these data. However, our tests on mock data indicate that we would have found evidence in favour of the H16 sub-halo with a log Bayes factor of 29 (equivalent to 7.6$\sigma$ significance) if it were present, a higher significance than the $6\sigma$ found by H16. 

I16 used a PL smooth model and a sub-halo parameterised as a homogeneous spherical clump of uniform density with a negative density shell localised to the southern point of the eastern arc. They inferred this model by fitting a PL and external shear to lensed source positions measured from a deconvolved image (made by tapering the data to an effective synthesised beam of 0.1\,arcsec) and adopting a trial-and-error approach. While the constrained image position is consistent with a positive multipole convergence feature (Fig.~\ref{fig:reconst}), it is very plausible that the differences in our findings are because of the different model assumptions and the number of degrees of freedom available using the full surface brightness of the lensed emission. For example, \cite{Powell:2022} noted significant differences in the lens model inferred for the system MG~J0751+2716 using a pixellated source reconstruction compared to a previous analysis using point-like image positions \citep{Spingola:2018}. Furthermore, recent work by \cite{Cohen:2024} show that multipole perturbations with amplitudes of 1~percent (consistent with our best-fit multipole amplitudes; see also \citealt{Stacey:2024}) can perfectly fit quasar image positions and flux ratios, and \cite{O'Riordan:2024} showed that multipoles, if not included in the model, can readily lead to artificial sub-halo constraints. 

On the other hand, H16 performed a full forward modelling of the source surface brightness using 40\,\% of the visibility data. We ruled out that the treatment of phase errors led to different results; evidence in favour of a sub-halo did not significantly change for mock data with random phase errors. Another key difference with our analysis is how the source is constructed. We used a Delaunay source grid adapted to the lensing magnification, but H16 used a regular source grid with a pixel size of 10~mas and a different surface brightness prior. \cite{Nightingale:2015} found that a regular source grid can induce systematic biases in the inferred lens model parameters. We did not test whether a 10~mas regular source grid sufficiently samples the source structure of SDP\,81, but \citeauthor{Nightingale:2015} found that a finer grid would not resolve these systematic biases. While we found that a 10\,mas regular source grid led to inconsistent inferred lens parameters for these data, it did not produce evidence in favour of a sub-halo in any location, so we cannot ascribe the H16 sub-halo inference to regular grid bias.

We inferred a PL model consistent with those inferred with previous work (e.g. \citealt{Dye:2015}), but different multipole coefficients from those inferred by H16. We find that our multipole coefficients are more consistent with the data, and the H16 coefficients are strongly disfavoured. Furthermore, the multipole structure was previously modelled and analysed by
\cite{Stacey:2024}, who found a strikingly similar structure in the stellar isophotes and mass isodensity shapes despite the stellar light not having been used as a prior for the lens model. 

We find that the most plausible origin of the H16 sub-halo is these inferred smooth model parameters. Their different multipole coefficients suggest that the statistical analysis performed by H16 did not identify the global minimum in the smooth model posterior. This is the only scenario we tested where the H16 sub-halo was favoured relative to its corresponding smooth model. In this scenario, the positive inference of a sub-halo is produced as a result of a poor smooth model fit: the additional degrees of freedom afforded by the sub-halo component act to reduce residuals.

Notably, this scenario we ascribe to the H16 sub-halo inference, is not the same problem discussed in Section~\ref{section:degeneracy} concerning the similar lensing signatures of multipoles and massive sub-haloes. Rather, it exemplifies how the inference of low-mass haloes requires a good smooth model fit. A poor smooth model fit can be characterised by the level of residuals in multiple lensed images For example, \cite{Nightingale:2022} set a criterion for the suitability of a lens for substructure analysis based on the level of smooth model residuals: they assessed that $>3\sigma$ residuals over multiple lensed images indicated a poor smooth model fit, such that inclusion of a sub-halo component in the lens model would not produce a reliable inference. In general, a poor smooth model fit may indicate non-localised complexity in the mass distribution beyond what can be described by PL+MP (e.g. see \citealt{Powell:2022}), but we find no need for complexity beyond PL+MP to fit these data.


\section{Conclusions}
\label{section:conclusions}

The `detection', or lack thereof, of low-mass dark matter haloes via gravitational lensing is a practical approach to testing dark matter models. Several recent works have explored systematic biases in inferring the presence of low-mass haloes, such as the complexity of the primary lens mass distribution
\citep{Nightingale:2022,O'Riordan:2024} and source resolution \citep{Vernardos:2022,Nightingale:2022,Ballard:2024,Minor:2025,Ephremidze:2025}. It is nonetheless essential that reported detections are independently verified.

Here, we have been unable to reproduce the inference of a sub-halo by \cite{Hezaveh:2016} in ALMA data of SDP\,81. Using mock data, we have confirmed that we would have found evidence in favour of that sub-halo should it exist with the prescribed properties. After testing several systematics (smooth model, source gridding, phase errors), we find that the most likely cause of this spurious inference was a poorly fitting multipole component, suggesting that their model inference did not find the global minimum in the parameter space. Another sub-halo reported by \cite{Inoue:2016} can be explained by several limitations in the earlier methodology and the presence of angular structure in the lens. Therefore, we conclude that these previous results are not robust.

We find that the presence of angular structure in the lens described by multipoles of 3rd and 4th order (see also \citealt{Stacey:2024}) is sufficient to explain the data, and we find no evidence in favour of any individual sub-halo. However, we find agreement with recent works indicating degeneracy between massive sub-haloes and multipoles \citep{O'Riordan:2024}. We suggest that future searches for low-mass perturbers via lensed arcs must allow for complex angular structure to avoid false inferences, which may be achieved by exploiting the full surface brightness of the lensed emission. The goodness of fit of this smooth model is an essential prerequisite to sub-halo inference, which may be assessed via the significance and distribution of image-plane residuals (e.g. \citealt{Nightingale:2022}).

Nevertheless, here, we have shown the ability to differentiate between models including a $\approx10^9\,{\rm M_\odot}$ sub-halo, demonstrating that ALMA data of this quality are useful to test the prevalence of sub-haloes in the $<10^{10}\,{\rm M_\odot}$ mass range (see also \citealt{Despali:2022}). In this case, the inferred multipole coefficients are largely stable in the presence of sub-haloes. Forthcoming analysis from 10 strongly lensed dusty star-forming galaxies observed with ALMA at similar angular resolution will give insight into the generalisability of this result.

\bigskip

\begin{acknowledgements}
We thank Yashar Hezaveh and Luke Maud for useful discussions. H.\,R.\,S.,  D.\,M.\,P. and S.\,V. have received funding from the European Research Council (ERC) under the European Union’s Horizon 2020 research and innovation programme (grant agreement No 758853). S.\,V. thanks the Max Planck Society for support through a Max Planck Lise Meitner Group. This research was carried out on the High-Performance Computing resources of the Freya cluster at the Max Planck Computing and Data Facility (MPCDF) in Garching, operated by the Max Planck Society (MPG). JPM and DW acknowledge support from the Netherlands Organization for Scientific Research (NWO) (Project No. 629.001.023) and the Chinese Academy of Sciences (CAS) (Project No. 114A11KYSB20170054). This work is based on the research supported in part by the National Research Foundation of South Africa (Grant Number: 128943). The modified bibliography style is courtesy of Chentao Yang (\url{https://github.com/yangcht/AA-bibstyle-with-hyperlink}).
This research used Astropy, SciPy, NumPy, and Matplotlib packages for Python \citep{Astropy:2013,Astropy:2018,Virtanen:2020,Harris:2020,Hunter:2007}.  We made use of ALMA data with project code 2011.0.00016.SV. ALMA is a partnership of ESO (representing its member states), NSF (USA) and NINS (Japan), together with NRC (Canada), MOST and ASIAA (Taiwan), and KASI (Republic of Korea), in cooperation with the Republic of Chile. The Joint ALMA Observatory is operated by ESO, AUI/NRAO and NAOJ.
\end{acknowledgements}



\bibliographystyle{aa_url}
\bibliography{references} 


\appendix
\renewcommand{\thefigure}{A\arabic{figure}}

\begin{figure*}
    \centering
    \includegraphics[width=\textwidth]{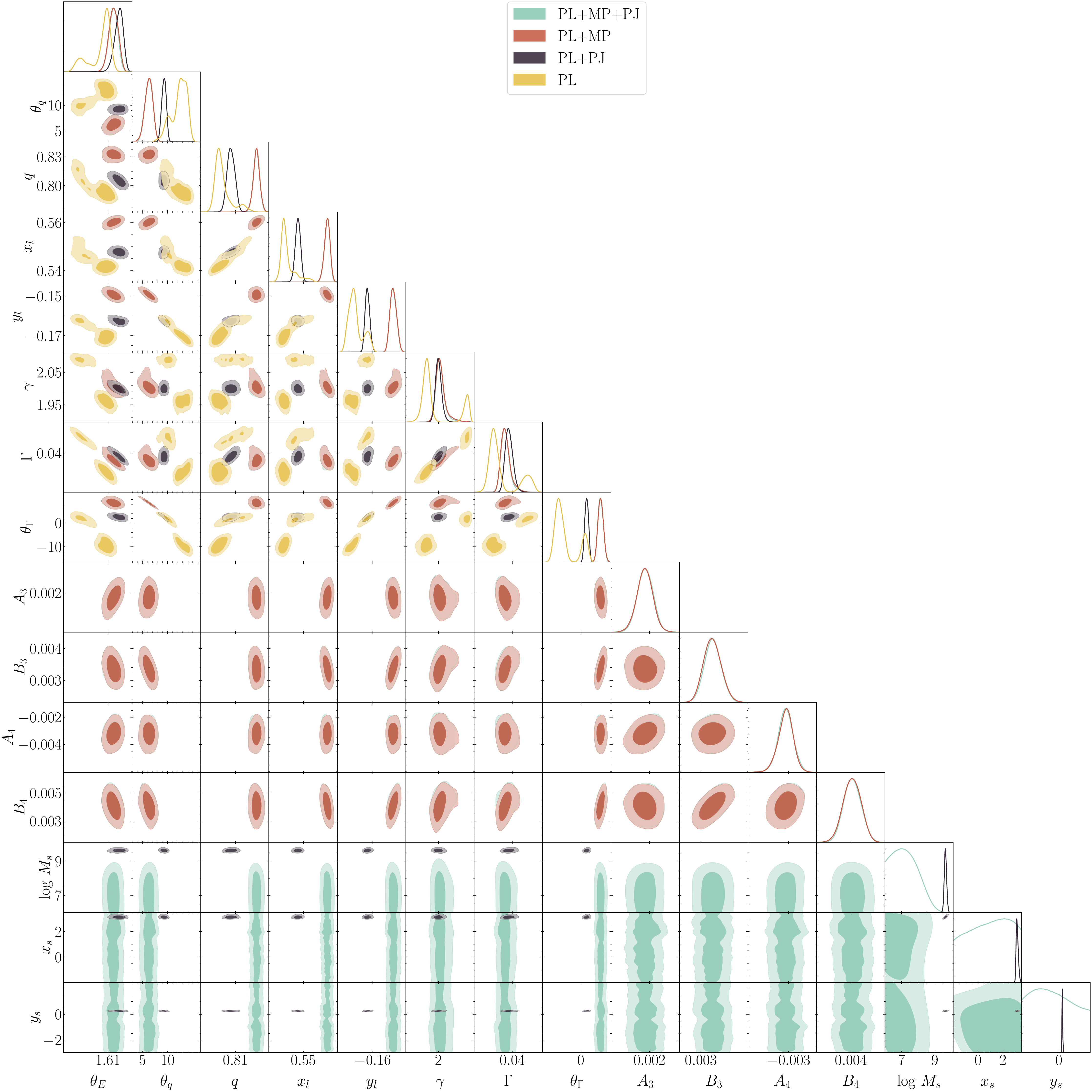}
    \caption{Two-dimensional posterior probability distributions of parameters of four models fit to the band 7 data of SDP\,81. Contours show 1$\sigma$ and 2$\sigma$ volumes of the posterior.}
    \label{fig:cornerplot}
\end{figure*}

\renewcommand{\thetable}{A\arabic{table}}

\begin{table*}
    \caption{ Lens model parameters for the band 7 data and their $1\sigma$ uncertainties (extended version of Table~\ref{table:evidence}). All positions are given relative to the observation phase centre. The log Bayes factor ($\ln\mathcal{K}$) is relative to the model with the highest evidence. The $M_{s}$ definition is described in Eq.~\ref{eq:MPJ}. Positions are given relative to the phase centre (J2000 09:03:11.61 +00:39:06.70) and angles are defined east of north. Parameter definitions follow the Code-independent Organised Lens Standard \citep{Galan:2023}.}
    \setlength{\tabcolsep}{3.5pt}
    \renewcommand*{\arraystretch}{1.12}
    \begin{tabularx}{\textwidth}{ l | c c c c c c} \hline 
                                         & PL            & PL+MP         & PL+PJ         & PL+MP+PJ      &  PL+MP+PJ (p) &   PL+MP+PJ (p,m)   \\  \hline 
    $x_l$ ($''$)                         & $0.542\pm0.001$ & $0.560\pm0.001$ & $0.548\pm0.001$ & $0.560\pm0.001$ & $0.560\pm0.001$ & $0.561\pm0.001$ \\ 
    $y_l$ ($''$)                         & $-0.170\pm0.003$ & $-0.149\pm0.002$ & $-0.163\pm0.001$ & $-0.149\pm0.002$ & $-0.149\pm0.002$ & $-0.144\pm0.001$ \\ 
    $\theta_E$ ($''$)                    & $1.609\pm0.004$ & $1.611\pm0.001$ & $1.613\pm0.001$ & $1.611\pm0.001$ & $1.611\pm0.001$ & $1.606\pm0.001$ \\ 
    $q$                                  & $0.794\pm0.005$ & $0.832\pm0.003$ & $0.805\pm0.004$ & $0.832\pm0.003$ & $0.832\pm0.003$ & $0.812\pm0.004$ \\ 
    $\theta_q$ ($^\circ$)                & $13\pm2$ & $6\pm1$ & $9.3\pm0.5$& $6\pm1$ & $6\pm1$ & $2\pm1$ \\ 
    $\gamma$                             & $1.97\pm0.01$ & $2.00\pm0.01$ & $2.00\pm0.01$ & $2.01\pm0.02$ & $2.01\pm0.02$ & $2.08\pm0.01$ \\ 
    $\Gamma$                             & $0.030\pm0.001$ & $0.037\pm0.002$ & $0.038\pm0.002$ & $0.037\pm0.002$ & $0.037\pm0.002$ & $0.044\pm0.002$ \\ 
    $\theta_\Gamma$ ($^\circ$)           & $-9\pm2$ & $9\pm1$ & $3\pm1$& $9\pm1$ & $8\pm1$ & $11\pm1$ \\ 
    $A_3$                                & -             & $0.0018\pm0.0004$ & -             & $0.0018\pm0.0004$ & $0.0018\pm0.0004$ & $0.0014\pm0.0003$ \\ 
    $B_3$                                & -             & $0.0034\pm0.0003$ & -             & $0.0034\pm0.0003$ & $0.0034\pm0.0003$ & $0.0037\pm0.0002$ \\ 
    $A_4$                                & -             & $-0.0032\pm0.0006$ & -            & $-0.0032\pm0.0006$ & $-0.0032\pm0.0006$ & $0.0000\pm0.0008$ \\ 
    $B_4$                                & -             & $0.0041\pm0.0007$ & -             & $0.0041\pm0.0006$ & $0.0039\pm0.0007$ & $-0.0045\pm0.0005$ \\ 
    $x_{\rm s}$ ($''$)                 & -             & -                & $3.1\pm0.1$& $0.9\pm1.5$ & - & - \\ 
    $y_{\rm s}$ ($''$)                 & -             & -                & $0.26\pm0.02$ & $-0.4\pm1.5$ & - & -\\ 
    $\log\,(M_{s}/{\rm M_{\odot}})$ & -         & -             & $9.6\pm0.1$ & $<8.1$ & $<7.8$ & - \\ \hline
    $\ln\mathcal{K}$                        & $-28$     & $0$        & $-6$     & $0$  & $0$ & $-13$      \\  \hline 
    \end{tabularx} 
    \label{table:evidence_ext}
\end{table*}

\begin{table*}
    \caption{ Lens model parameters for the band 7 data and their $1\sigma$ uncertainties, using a regular source grid. Parameter definitions are as in Table~\ref{table:evidence_ext}. $^{\ddagger}$ indicates that this parameter is not constrained (it is constrained only by the prior).}
    \setlength{\tabcolsep}{3.5pt}
    \renewcommand*{\arraystretch}{1.11}
    \begin{tabularx}{0.87\textwidth}{ l | c c c c c } \hline 
                                         & PL            & PL+MP                    & PL+MP+PJ          &  PL+MP+PJ (p)     &   PL+MP+PJ\,(p,m)   \\  \hline 
    $x_l$ ($''$)                         & $0.541\pm0.001$ & $0.569\pm0.002$        & $0.552\pm0.001$   & $0.560\pm0.002$   & $0.547\pm0.002$ \\ 
    $y_l$ ($''$)                         & $-0.170\pm0.001$ & $-0.173\pm0.001$      & $-0.163\pm0.001$  & $-0.163\pm0.001$  & $-0.158\pm0.001$\\ 
    $\theta_E$ ($''$)                    & $1.611\pm0.001$ & $1.615\pm0.001$        & $1.621\pm0.001$   & $1.621\pm0.001$   & $1.622\pm0.002$ \\ 
    $q$                                  & $0.793\pm0.004$ & $0.825\pm0.003$        & $0.796\pm0.004$   & $0.795\pm0.005$   & $0.774\pm0.006$ \\ 
    $\theta_q$ ($^\circ$)                & $13\pm1$ & $6\pm1$ & $16\pm1$                                & $9\pm1$           & $7\pm1$  \\ 
    $\gamma$                             & $1.94\pm0.01^{\ddagger}$ & $1.902\pm0.002^{\ddagger}$ & $1.913\pm0.004^{\ddagger}$ & $1.912\pm0.005^{\ddagger}$ & $1.95\pm0.02^{\ddagger}$  \\ 
    $\Gamma$                             & $0.029\pm0.001$ & $0.030\pm0.001$        & $0.016\pm0.001$   & $0.016\pm0.002$   & $0.015\pm0.003$  \\ 
    $\theta_\Gamma$ ($^\circ$)           & $-10\pm1$ & $-16\pm1$                    & $-7\pm2$          & $-7\pm3$          & $-11\pm4$    \\ 
    $A_3$                                & -             & $0.0023\pm0.0003$        & $0.0036\pm0.0003$ & $0.0035\pm0.0004$   & $0.0043\pm0.0006$ \\ 
    $B_3$                                & -             & $-0.0002\pm0.0003$       & $0.0013\pm0.0002$ & $0.0013\pm0.0002$   & $0.0001\pm0.0003$ \\ 
    $A_4$                                & -             & $-0.007\pm0.001$         & $-0.0019\pm0.0004$ & $-0.0021\pm0.0005$ & $0.0001\pm0.0001$  \\ 
    $B_4$                                & -             & $-0.0056\pm0.0006$       & $-0.0022\pm0.0004$ & $-0.0022\pm0.0005$ & $0.0047\pm0.0007$  \\ 
    $x_{\rm s}$ ($''$)                 & -             & -                          & $1.1\pm1.5$        & -                & - \\ 
    $y_{\rm s}$ ($''$)                 & -             & -                          & $-0.3\pm1.5$       & -                 & - \\ 
    $\log\,(M_{s}/{\rm M_{\odot}})$ & -         & -                                 & $9.6\pm0.1$        & $<7.6$           & - \\ \hline
    $\ln\mathcal{K}$                        & $-32$     & $0$                       & $-5$               & $-6$              & $-70$      \\  \hline 
    \end{tabularx} 
    \label{table:evidence_ext_reg}
\end{table*}

\begin{table*}
    \caption{ Lens model parameters and their $1\sigma$ uncertainties for mock band 7 data with ground truth that is our maximum {\it a posteriori} PL+MP model inferred for the real band 7 data. Parameter definitions are as in Table~\ref{table:evidence_ext}. }
    \setlength{\tabcolsep}{3.5pt}
    \renewcommand*{\arraystretch}{1.11}
    \begin{tabularx}{0.56\textwidth}{ l | c c c } \hline 
                                         & PL               & PL+MP                     & PL+PJ           \\  \hline 
    $x_l$ ($''$)                         & $0.562\pm0.001$  & $0.569\pm0.002$           & $0.568\pm0.001$   \\ 
    $y_l$ ($''$)                         & $-0.152\pm0.002$ & $-0.144\pm0.002$           & $-0.142\pm0.002$   \\ 
    $\theta_E$ ($''$)                    & $1.604\pm0.002$  & $1.612\pm0.001$           & $1.614\pm0.001$   \\ 
    $q$                                  & $0.829\pm0.006$  & $0.835\pm0.005$           & $0.814\pm0.004$   \\ 
    $\theta_q$ ($^\circ$)                & $9\pm1$          & $7\pm1$                   & $5\pm1$         \\ 
    $\gamma$                             & $2.07\pm0.01$    & $2.01\pm0.02$             & $2.03\pm0.02$ \\ 
    $\Gamma$                             & $0.048\pm0.003$  & $0.037\pm0.003$           & $0.034\pm0.003$   \\ 
    $\theta_\Gamma$ ($^\circ$)           & $4\pm1$         & $8\pm2$                    & $3\pm2$          \\ 
    $A_3$                                & -                & $0.0019\pm0.0004$        & -  \\ 
    $B_3$                                & -                & $0.0035\pm0.0004$        & -  \\ 
    $A_4$                                & -                & $-0.003\pm0.001$          & -  \\ 
    $B_4$                                & -                & $0.003\pm0.001$        & -  \\ 
    $x_{\rm s}$ ($''$)                  & -                  & -                          & $-1.8\pm0.1$        \\ 
    $y_{\rm s}$ ($''$)                  & -                  & -                          & $1.8\pm0.2$        \\ 
    $\log\,(M_{s}/{\rm M_{\odot}})$      & -                 & -                          & $10.1\pm0.1$        \\ \hline
    $\ln\mathcal{K}$                     & $-27$            & $0$                     & $-3$              \\  \hline 
    \end{tabularx} 
    \label{table:evidence_ext_mp} 
\end{table*}

\begin{table*}
    \caption{ Lens model parameters and their $1\sigma$ uncertainties for mock band 7 data with ground truth PL+MP+PJ\,(p,m), where PL+MP is our maximum {\it a posteriori} model inferred for the real band 7 data. Parameter definitions are as in Table~\ref{table:evidence_ext}. }
    \setlength{\tabcolsep}{3.5pt}
    \renewcommand*{\arraystretch}{1.11}
    \begin{tabularx}{0.58\textwidth}{ l | c c c } \hline 
                                         & PL               & PL+MP                     & PL+MP+PJ           \\  \hline 
    $x_l$ ($''$)                         & $0.562\pm0.002$  & $0.559\pm0.002$           & $0.560\pm0.001$   \\ 
    $y_l$ ($''$)                         & $-0.145\pm0.002$ & $-0.148\pm0.002$          & $-0.149\pm0.001$  \\ 
    $\theta_E$ ($''$)                    & $1.611\pm0.001$  & $1.611\pm0.002$           & $1.611\pm0.001$   \\ 
    $q$                                  & $0.832\pm0.004$  & $0.815\pm0.001$           & $0.806\pm0.005$   \\ 
    $\theta_q$ ($^\circ$)                & $4\pm1$          & $7\pm1$                   & $7\pm1$         \\ 
    $\gamma$                             & $2.04\pm0.02$    & $2.07\pm0.02$             & $2.09\pm0.01$ \\ 
    $\Gamma$                             & $0.043\pm0.003$  & $0.042\pm0.002$           & $0.041\pm0.001$   \\ 
    $\theta_\Gamma$ ($^\circ$)           & $11\pm1$         & $6\pm1$                    & $6\pm1$          \\ 
    $A_3$                                & -                & $-0.0005\pm0.0006$        & $-0.0004\pm0.0003$  \\ 
    $B_3$                                & -                & $-0.0022\pm0.0004$        & $-0.021\pm0.0003$  \\ 
    $A_4$                                & -                & $-0.002\pm0.001$          & $-0.0024\pm0.0001$  \\ 
    $B_4$                                & -                & $-0.0023\pm0.0007$        & $-0.0032\pm0.0006$  \\ 
    $x_{\rm s}$ ($''$)                 & -                  & -                          & $-0.79\pm0.02$        \\ 
    $y_{\rm s}$ ($''$)                 & -                  & -                          & $-0.67\pm0.02$        \\ 
    $\log\,(M_{s}/{\rm M_{\odot}})$     & -                 & -                          & $8.4\pm0.1$        \\ \hline
    $\ln\mathcal{K}$                     & $-42$            & $-29$                     & $0$              \\  \hline 
    \end{tabularx} 
    \label{table:evidence_ext_pj} 
\end{table*}

\end{document}